\documentclass[11pt,twoside]{article}

\usepackage{asp2014}

\aspSuppressVolSlug
\resetcounters

\begin{document}

	\title{PolarVis: Towards Web-based Polarimetric Analysis}
	
	\author{L. A. L. Andati,$^1$ O. M. Smirnov,$^{1,2}$ S. Makhathini,$^3$, and L. M. Sebokolodi$^1$}
	\affil{$^1$Rhodes University, Makhanda, South Africa; \email{andatilexy@gmail.com}}
	\affil{$^2$South African Radio Astronomy Observatory, Capetown, Western Cape, South Africa.}
	\affil{$^3$University of the Witwatersrand, Johannesburg, Gauteng, South Africa.}

	\paperauthor{L.~A.~L.~Andati}{andatilexy@gmail.com}{0000-0002-7629-4370}{Centre for Radio Astronomy Techniques and Technologies, Rhodes University}{Department of Physics and Electronics}{Makhanda}{Eastern Cape}{6139}{South Africa}
	\paperauthor{O.~M.~Smirnov}{Author2Email@email.edu}{ORCID_Or_Blank}{South African Radio Astronomy Observatory}{Author2 Department}{Capetown}{Western Cape}{7925}{South Africa}
	\paperauthor{S.~Makhathini}{Author3Email@email.edu}{ORCID_Or_Blank}{Author3 Institution}{Author3 Department}{City}{State/Province}{Postal Code}{Country}
	\paperauthor{L.~M.~Sebokolodi}{andatilexy@gmail.com}{0000-0002-7629-4370}{Centre for Radio Astronomy Techniques and Technologies, Rhodes University}{Department of Physics and Electronics}{Makhanda}{Eastern Cape}{6139}{South Africa}

	
	
	\begin{abstract}
     Astronomers performing polarimetric analysis on astronomical images often have to manually identify locations on their objects of interest, such as galaxies, which exhibit the influence of magnetic forces due to interaction with their environments or inherent processes. These locations are known as Lines of Sight (LoS). Analysing the various lines of sight can provide insight into the electromagnetic nature of the astrophysical object in question and its surroundings. For each LoS, astronomers generate diagnostic plots to map out the variation of the corresponding electromagnetic field, such as those of fractional polarisation and Faraday spectra. However, associating the different LoS diagnostic plots to their positions on an astronomical image requires alternating between the plots and the images. As a result, determining whether the location of the LoS influences its magnetic field variation by analysing its diagnostic plots becomes arduous due to the absence of a direct way of linking the two. \texttt{PolarVis} is an effort towards allowing an almost instant view of the interactive diagnostic plots corresponding to a given line of sight at the click of a button on that line of sight on the image, using an interactive web-based FITS viewer -JS9.
	\end{abstract}
	
	

\section{Introduction}
An electromagnetic (EM) wave results from oscillating electric and magnetic fields orthogonal to their direction of propagation. That of the electric field determines the wave's state of polarisation. In radio astronomy, we define the emission properties of the EM wave in terms of Stokes parameters: $I$, $Q$, $U$, and $V$. $I$ represents the total intensity, $Q$ and $U$ are linear polarisation, and $V$ is circular polarisation. The total linear polarised intensity is defined as:

\begin{equation}\label{eq1}
P (\lambda) = Q + iU.
\end{equation}

In astronomy, polarisation data is a useful diagnostic that can reveal information about an astronomical source and the media through which its radiation propagates. Its analysis is used extensively in astronomy to study the strength and structure of the magnetic fields permeating the cosmos through the study of the Faraday rotation effect. Faraday Rotation occurs when a plane of polarisation of a linearly polarised wave gets rotated as it passes through a magnetised plasma - such as the interstellar, intergalactic, intracluster gas.

In addition to Eq. \ref{eq1}, astronomers commonly use a technique known as RM-synthesis to study cosmic magnetic fields. This technique Fourier transforms Eq. \ref{eq1}:
\begin{equation}
F(\phi) = \int_{-\infty}^{\infty} P (\lambda) e^{-2i\phi \lambda^2} d\lambda^2,
\end{equation}
where $\phi$ is the Faraday depth:
\begin{equation}
\phi = 812 \int_{\text{source}}^{\text{observer}} n_e B_{\text{LoS}} dl, 
\end{equation}
$F(\phi)$ is polarised emission at $\phi$ also known as Faraday spectrum, $\lambda^2$ observing wavelength, $n_e$ [cm$^{-3}$] is the electron density of the plasma, and $B_{\text{LoS}}$ [$\mu$G] is the lines of sight (LoS) component of the magnetic field (see \citet{burn1966depolarization}  and \citet{brentjens2005faraday} for further details). 

Recent polarimetry studies reveal remarkable complexity in the polarisation structures across the lobes of radio galaxies. Additionally, the amount of data involved is very large -- e.g  \citet{sebokolodi2020wideband} had $2000$ lines of sight across Cygnus A radio lobes. In that study, everything had to be done manually e.g classification of the different polarisation structures -- a very painstaking process.

\section{PolarVis}
\texttt{PolarVis}\footnote{Visit \url{https://cygnus.ratt.center/cygnus} to view and interact with \texttt{PolarVis}} is a web-based visualisation tool useful for inspecting polarisation spectra for various LoS on astronomical data images (FITS images). It is built on the  JS9\footnote{\url{https://js9.si.edu/}} FITS image viewer and \texttt{Bokeh}\footnote{\url{https://bokeh.org/}}, a \texttt{Python} based visualisation library. We use JS9 as it exposes simple, event hooks which allow developers to attach custom reactions triggered by actions on the web browser, such as clicking. The hooks can be attached to an entire FITS image or particular regions within a single image. \texttt{PolarVis} takes advantage of JS9's event-driven hooks to provide an interactive interface over each of the various LoS. A line of sight is interpreted as a region which when clicked, automatically generates its corresponding Faraday and fractional polarisation spectra. Each LoS is green in colour as shown in Fig. \ref{polarvis_layout}.

\articlefigure{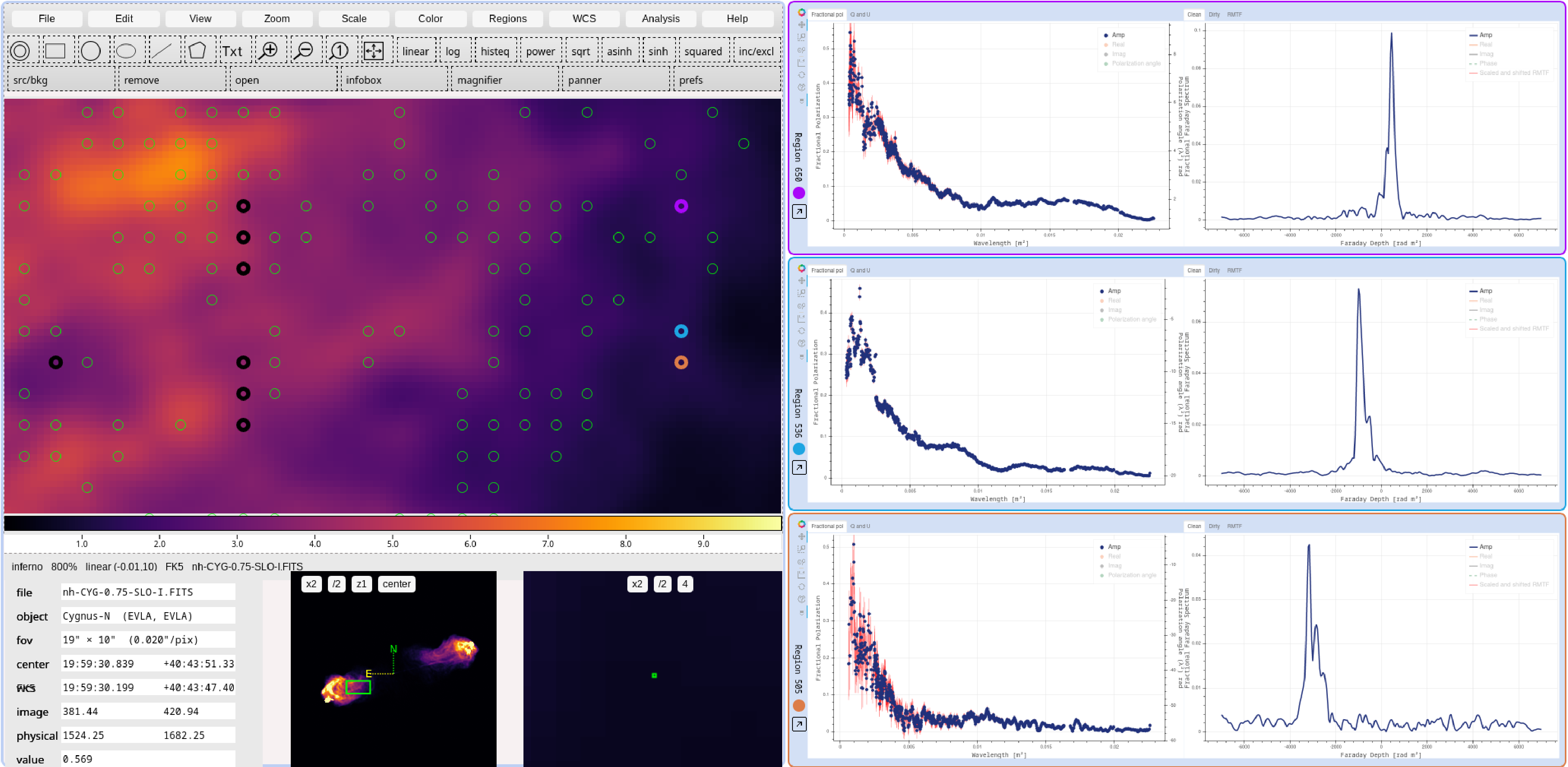}{polarvis_layout}{\texttt{PolarVis} has a two-column layout where the left-hand side contains the FITS image viewer (JS9), and the right-hand side the plots of respective LoSs. LoSs are denoted by the green, thin circles on the FITS image. Older selected LoS are denoted by thicker circles, with the most recent selections coloured purple, blue and orange (on the rightmost side of the FITS image) and the rest coloured black (these are the ``breadcrumbs''). It is possible to identify the three most recent selections as their colours correspond to the colours on the leftmost side of the plots}

\texttt{PolarVis} leaves ``breadcrumbs'' showing which LoSs have been previously clicked by changing the region's colour, as illustrated on the left-hand side of Fig \ref{polarvis_layout}. For example, the first clicked region is purple, while the second one is orange and the third blue. These three colours always highlight the three most recently selected LoS. If there have been more than three previous selections, the oldest selections then change to the colour black. The three aforementioned colours serve a further purpose of indicating the corresponding plot on display as the plots are marked with the same colour.

\articlefigure{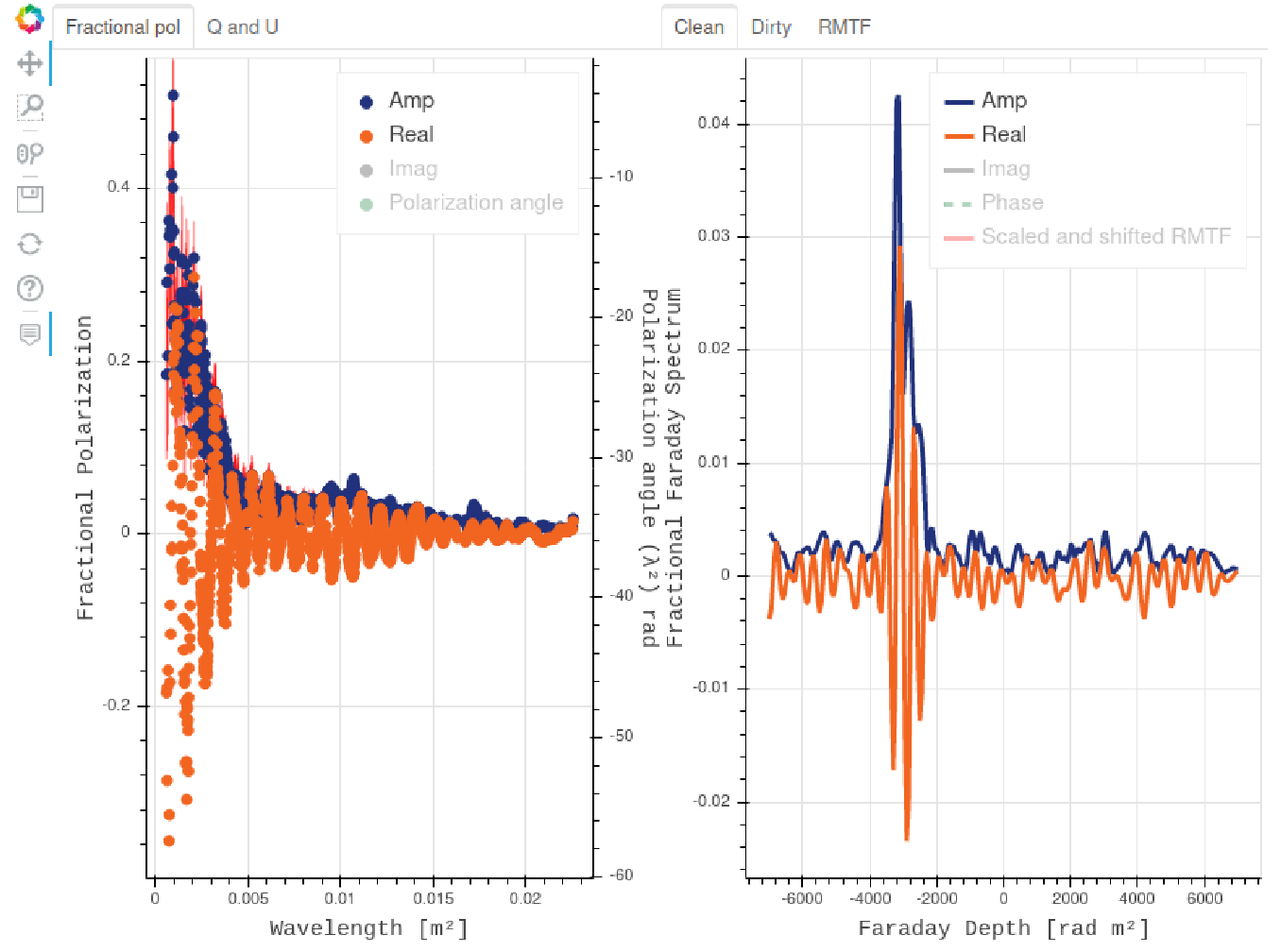}{plot_layout}{An example of an LoS spectra plots. Interactivity tools are on the leftmost side of the plot, while tabs atop the plots allow switching between various plots. The legends are also clickable and can make some plots visible or hidden.}

\clearpage

Plots of the various spectra are also interactive. They can be enlarged, zoomed, panned, made visible or hidden using an array of tools placed towards the left-hand side of the plot, and by clicking on the plot legends. Fig. \ref{plot_layout} demonstrates a sample plot. Moreover, it is possible to select between different plots, by clicking on tabs at the top of each plot. This kind of set-up enables the collection of more plots corresponding to a single LoS in a single, accessible place.

In a more ``traditional'' setup FITS images and FITS viewers resided in the same local machine, translating to a significantly reduced latency between opening a FITS image and seeing it. Web-based FITS visualisation differs, however, because the image must be transported over a network for a user to see. If we consider a fiducial FITS image of $\approx$ 67MB in size, transferring this amount of data over a network would and eventually loading up a web page and displaying the image would take a considerable amount of time (up to minutes) leading to a sluggish and unpleasant user experience.
\texttt{PolarVis} mitigates this latency by leveraging JS9's ability to display smaller versions of the original FITS image, thereby reducing the bandwidth used, and the time taken to load up the image. Currently, an image of 67MB in size requires just $\approx$4MB to load up and view, which is a significant size reduction.

\section{Conclusion}
We have presented a web-based polarimetric analysis tool, \texttt{PolarVis}, built on JS9's existing and powerful infrastructure. This tool matches various interesting pre-selected regions known as Lines of Sight (LoS) to their corresponding pre-generated spectral plots for ease of viewing and comparison by the user. The plots are accessible by clicking on the desired LoS and are inherently also interactive such that they allow zooming, panning and can be selected to be made visible or hidden. \texttt{PolarVis} also uses minimal bandwidth to achieve its goal. Visit \url{https://cygnus.ratt.center/} to interact with it.

\acknowledgements This research was funded by the South African Radio Astronomy Observatory (SARAO) through the Centre for Radio Astronomy Techniques and Technologies (RATT). Additional thanks to Lerato for lending us data.

\bibliography{refs}  


\end{document}